\documentclass[aps,twocolumn,prl,showpacs,floatfix,superscriptaddress]{revtex4}
\usepackage{amsfonts}
\usepackage{graphicx}
\usepackage{amsmath}
\usepackage{amssymb}

\input{tcilatex}

\begin{document}

%
%
%



\title{Fractionally Quantized Hall Effect:
Liquid-Crystal Ground-State and it Excitations}

\title{Fractionally Quantized Hall Effect:
Liquid-Crystal Ground-State and it Excitations}

\author{O. G. Balev}
\email[Electronic address: ]{ogbalev@ufam.edu.br}
\affiliation{Departamento de F\'{\i}sica, Universidade
Federal do Amazonas, Manaus, AM 69077-000, Brazil}

\begin{abstract}
It is shown that the ground-state and the lowest excited-states of
two-dimensional electron system (2DES), with ion jellium background,
correspond to partial crystal-like (with the period
$L_{x}^{\square}=\sqrt{2 m \pi} \ell_{0}$,
along $x$; $\ell_{0}$ is the magnetic length)
correlation order among $N$ electrons of the main
region (MR; $L_{x} \times L_{y}$, $L_{x,y} \rightarrow \infty$).
Many-body variational ground-state wave function of 2DES
is presented at
the fractional and the integral filling factors $\nu=1/m$;
$m=2\ell+1$ and $\ell=0, \; 1, \; 2,\ldots$.
The ground-state manifests the broken symmetry liquid-crystal
state with 2DES density that is periodic along the $y-$ direction, with the
period $L_{x}^{\square}/m$, and independent of $x$.
At $m=3, \;5$, the ground-state has
essentially lower energy per electron than the Laughlin, uniform liquid,
ground-state; the same holds at $m=1$.
At $m \geq 3$, the compound form of the many-body ground-state wave
function leads to the compound structure of each electron, even
within the main strip (MS; $L_{x}^{\square} \times L_{y}$).
These compound electrons play important role in the properties of the
many-body excited-states. Obtained compound exciton (without the change of
spin of the excited compound electron) and compound spin-exciton (with the
change of spin of the excited compound electron) states show finite
excitation gaps, for $m=1, \; 3, \; 5$. The excited compound electron
(hole) is composed, within MS, from $m$ strongly correlated
quasielectrons (quasiholes) of the total charge $e/m$ ($-e/m$) each;
fractionally quantized at $m \geq 3$.
These $m$ quasielectrons (quasiholes) are periodically ``repeated''
outside MS, along $x$, with the period $L_{x}^{\square}$.
The activation gap
is obtained: it is given by the excitation
gap of relevant compound exciton, at $m \geq 3$, and by the gap of pertinent
compound spin-exciton, at $m=1$. Quantized Hall conductance 
$\sigma_{H}=e^{2}/(2 m \pi \hbar)$ is obtained; it is fractional at
$m \geq 3$.  The theory is in good agreement with experiments.

\end{abstract}

\date{October 5, 2007}

\maketitle

\section{Introduction. Theoretical framework}

To date understanding is that at the filling factor
$\nu=1/3$, $1/5$ the Laughlin
variational wave function \cite{laughlin1983} gives the best
known so far analytical approximation of exact many-body ground-state
wave function. At $\nu=1/3$, there were many attempts
to obtain lower energy for ground-state than Laughlin's
\cite{laughlin1983} incompressible liquid state, e.g., see references in
\cite{balev2006}. Here we will start with the
same many-electron Hamiltonian,
$\hat{H}(\textbf{r}_{1},\ldots,\textbf{r}_{N})$, for
2DES of $N$ electrons as in \cite{laughlin1983,morf1986},
only we adopt the Landau gauge for vector potencial
$\textbf{A}=-By\hat{\textbf{x}}$. We assume that $N$ electrons
are localized in MR of $(z=0)-$plane;
$N/L_{x}L_{y}=\nu/2\pi \ell_{0}^{2}$,
where $\nu=1/m$,
$\ell_{0}=\sqrt{\hbar c/|e|B}$.
As in \cite{laughlin1983,morf1986}, the classical
model of neutralizing ion jellium background (IJB) is used,
cf. \cite{balev2006}. Eigenstates
$\Psi(\textbf{r}_{1},\ldots,\textbf{r}_{N})$ of the Hamiltonian
$\hat{H}(\textbf{r}_{1},\ldots,\textbf{r}_{N})$ and their energies
are determined by
\begin{equation}
\hat{H}(\textbf{r}_{1},\ldots,\textbf{r}_{N})
\Psi(\textbf{r}_{1},\ldots,\textbf{r}_{N})=E_{N}
\Psi(\textbf{r}_{1},\ldots,\textbf{r}_{N}) .
\label{1}
\end{equation}%
We can assume that 2DES, with IJB, is located within the ribbon of
width $L_{y}$ bent into loop of radius $L_{x}/2\pi$.
Then Born-Carman periodic boundary conditions
$\textbf{r}_{i} \pm L_{x}\hat{\textbf{x}}=\textbf{r}_{i}$ are
holded, where $i=1,\dots,N$.

It is seen that the area of MR per electron, $L_{x}L_{y}/N=
(L_{x}^{\square})^{2}$, where $L_{x}^{\square}=\sqrt{2 m \pi}\ell_{0}$.
Then the strip of the width $L_{x}^{\square}$, along $x$-direction,
and of the length $L_{y}$ contains $\tilde{N}=L_{y}/L_{x}^{\square}$
of the (square) unit cells,
$L_{x}^{\square} \times L_{x}^{\square}$. The integer number
of such strips within MR is
given as $n_{xs}^{\max}=L_{x}/L_{x}^{\square}=N/\tilde{N}$;
for definiteness, odd (as $\tilde{N}$). Now we assume
that the ground-state and, at the least, the lowest excited-states of
Eq. (\ref{1}) correspond to partial crystal-like correlation order
among $N$ electrons of MR as
\begin{eqnarray}
&&\textbf{r}_{1+k \tilde{N}}=\textbf{r}_{1}+k L_{x}^{\square}
\hat{\textbf{x}}, \;\;\;
\textbf{r}_{2+k \tilde{N}}=\textbf{r}_{2}+k L_{x}^{\square}
\hat{\textbf{x}},
\notag \\
&& \cdots, \;\;\;
\textbf{r}_{\tilde{N}+k \tilde{N}}=
\textbf{r}_{\tilde{N}}+k L_{x}^{\square} \hat{\textbf{x}} ,
\label{2}
\end{eqnarray}%
where $k=0, \pm 1,\ldots,\pm (n_{xs}^{\max}-1)/2$.
Then Hamiltonian in Eq. (\ref{1}) becomes:
i) dependent only on $\tilde{N}$ $\mathbf{r}_{i}=(x_{i},y_{i})$
and ii) periodic, with period $L_{x}^{\square}$,
on any $x_{i}$; $i=1,\ldots,\tilde{N}$.
For the wave function in Eq. (\ref{1}): the property (i)
obviously holds and it is natural to assume that the property (ii)
is valid as well. Then
the study within MR of Eq. (\ref{1}), for 2DES of $N$ electrons,
can be reduced to the treatment of the
Schr\"{o}dinger equation for 2DES of $\tilde{N}$ ``compound'' electrons
within the main strip (MS) $L_{x}^{\square}n_{xs}^{MS}
>x_{i}>L_{x}^{\square}(n_{xs}^{MS}-1)$, cf. Ref. \cite{balev2006}, as
\begin{equation}
\hat{H}_{\tilde{N}}(\textbf{r}_{1},\ldots,\textbf{r}_{\tilde{N}})
\Psi_{\tilde{N}}=E_{\tilde{N}}
\Psi_{\tilde{N}}(\textbf{r}_{1},\ldots,\textbf{r}_{\tilde{N}}) ,
\label{3}
\end{equation}%
where, for definiteness, we assume $n_{xs}^{MS}=0$.
In Eq. (\ref{3})
\begin{equation}
\hat{H}_{\tilde{N}}=\hat{H}_{0}+V_{ee}+V_{eb}+V_{bb} ,
\label{4}
\end{equation}%
where the kinetic energy term
\begin{equation}
\hat{H}_{0}=\sum_{i=1}^{\tilde{N}}\hat{h}_{0i}=\frac{1}{2m^{\ast }}%
\sum_{i=1}^{\tilde{N}}[\hat{\mathbf{p}}_{i}-\frac{e}{c}\mathbf{A}(\mathbf{r}%
_{i})]^{2},
\label{5}
\end{equation}%
here $\hat{\mathbf{p}}=-i\hbar \mathbf{\nabla }$, $m^{\ast }$ the
electron effective mass.
In Eq. (\ref{4}) the 
electron-electron potential, cf. \cite{yoshioka1983}
\begin{eqnarray}
V_{ee} &=&\frac{1}{2}\sum_{i=1}^{\tilde{N}}\sum_{j=1,j\neq i}^{\tilde{N}%
}\sum_{k=-N_{C}}^{N_{C}}\frac{e^{2}}{\varepsilon |\mathbf{r}_{i}-\mathbf{r}%
_{j}-kL_{x}^{\square }\hat{\mathbf{x}}|}  \notag \\
&&+\tilde{N}\sum_{k=1}^{N_{C}}\frac{e^{2}}{\varepsilon L_{x}^{\square }k},
\label{6}
\end{eqnarray}%
physical results will not depend on $N_{C}\rightarrow \infty $.
The
electron-IJB interaction potential, cf. \cite{morf1986},
\begin{equation}
V_{eb}=-\sum_{i=1}^{\tilde{N}}\int_{MR}d\mathbf{R}
\frac{e^{2}n_{b}(\mathbf{R}%
)}{\varepsilon |\mathbf{r}_{i}-\mathbf{R}|},
\label{7}
\end{equation}%
and IJB-IJB interaction term
\begin{equation}
V_{bb}=\frac{1}{2}\int_{MS}d\mathbf{R}\int_{MR}d\mathbf{R}^{\prime }\frac{%
e^{2}n_{b}(\mathbf{R})n_{b}(\mathbf{R}^{\prime })}{\varepsilon |\mathbf{R}-%
\mathbf{R}^{\prime }|},  \label{8}
\end{equation}%
where the subscript MR (MS) shows that integration is carried out
over MR (MS); $n_{b}(\mathbf{R})=const=n_{b}$ and
$\int_{MS}d\mathbf{R}n_{b}=\tilde{N}$.
Notice $\langle \Psi _{\tilde{N}}| \Psi_{\tilde{N}}\rangle=1$, where
integrations are over the MS, and
$\epsilon=E_{\tilde{N}}/\tilde{N}=E_{N}/N$ is the total energy
per electron.

At $\nu=1/m$ we assume ground-state
$\Psi _{\tilde{N}}^{(m)}(\mathbf{r}_{1},\ldots ,\mathbf{r}_{\tilde{N}})$
of Eq. (\ref{3}) as \cite{balev2006}
\begin{equation}\Psi _{\tilde{N}}^{(m)}=
\sum_{n=-\ell }^{\ell }C_{n}(m)\Psi _{%
\tilde{N}}^{n,(m)}(\mathbf{r}_{1},\ldots , \mathbf{r}_{\tilde{N}}) ,
\label{9}
\end{equation}%
where $|C_{n}(m)|^{2}=1/m$ and
\begin{equation}
\Psi _{\tilde{N}}^{n,(m)}=\frac{1}{\sqrt{\tilde{N}!}}%
\begin{vmatrix}
\varphi _{k_{x1}^{(n)}}^{(m)}(\mathbf{r}_{1}) & \cdots & \varphi
_{k_{x1}^{(n)}}^{(m)}(\mathbf{r}_{\tilde{N}}) \\
\varphi _{k_{x2}^{(n)}}^{(m)}(\mathbf{r}_{1}) & \cdots & \varphi
_{k_{x2}^{(n)}}^{(m)}(\mathbf{r}_{\tilde{N}}) \\
\vdots & \ddots & \vdots \\
\varphi _{k_{x\tilde{N}}^{(n)}}^{(m)}(\mathbf{r}_{1}) & \cdots & \varphi
_{k_{x\tilde{N}}^{(n)}}^{(m)}(\mathbf{r}_{\tilde{N}})%
\end{vmatrix}%
,  \label{20a}
\end{equation}%
is
the Slater determinant
of single-electron wave functions (orthonormal within MS;
$y_{0}(k_{x})=\ell_{0}^{2}k_{x}$ )
\begin{equation}
\varphi _{k_{xi}^{(n)}}^{(m)}(\mathbf{r})=
e^{ik_{xi}^{(n)} x} \Psi_{0}(y-y_{0}(k_{xi}^{(n)})) /\sqrt{%
L_{x}^{\square}},
\label{10}
\end{equation}%
where $\Psi_{0}(y)$ is the harmonic oscillator function,
$i=1,2,\ldots,\tilde{N}$ is the number of a unit cell within MS,
\begin{equation}
k_{xi}^{(n)}=\frac{2\pi \; m}{L_{x}^{\square}} \left[%
n_{ys}^{(i)} + \frac{n}{m} \right] ,  \label{11}
\end{equation}%
here $n_{ys}^{(i)}=0, \pm 1, \dots,\pm(\tilde{N}-1)/2$,
the "set" number $n=0,\ldots,\pm \ell$;
$\langle \Psi _{\tilde{N}}^{k,(m)}|
\Psi _{\tilde{N}}^{n,(m)}\rangle =\delta _{k,n}$.
Notice, for the ground-state
Eq. (\ref{9}) electron density \cite{balev2006},
\begin{equation}
n(y)=\frac{m^{-1}}{2\pi \ell _{0}^{2}}[1+2\sum_{k=1}^{\infty }
e^{-\frac{\pi mk^{2}}{2}}\cos (\frac{\sqrt{2\pi m}}{\ell _{0}}ky)],
\label{12}
\end{equation}
depends only on $y$, periodically; $L_{x}^{\square}$ is the period.
It is readily seen that the periodicity property (ii) is valid as
for ground-state wave function, Eq. (\ref{9}), so for
$\Psi _{\tilde{N}}^{n,(m)}(\mathbf{r}_{1},\ldots ,\mathbf{r}_{\tilde{N}})$.

\section{Results and Discussion}

Using Eq. (\ref{4}) we obtain
the energy $\epsilon^{(m)}=\hbar \omega_{c}/2+
(e^{2}/\varepsilon \ell_{0}) U(m)$ of the ground-state
Eq. (\ref{9}) as $<\Psi _{\tilde{N}}^{(m)}|\hat{H}_{\tilde{N}}|
\Psi _{\tilde{N}}^{(m)}>/\tilde{N}$, where $U(m)$ is given by
analytical expressions,
deduced explicitly in
\cite{balev2006}. After simple numerical
calculations, we obtain: $U(3)=-0.42854$, $U(5)=-0.33885$,
and $U(1)=-0.66510$. So $U(3)$, $U(5)$, and $U(1)$ there are
substantially lower than pertinent total lowering at $\nu=1/3$,
$1/5$, and $1$ for the Laughlin's variational wave function
\cite{laughlin1983}: $-0.4156\pm 0.0012$ ($-0.410 \pm 0.001$
is obtained in Ref. \cite{morf1986}, by fixing relevant
numerical study of \cite{laughlin1983}), $-0.3340\pm0.0028$, and
$-\sqrt{\pi/8} \approx -0.6267$.

We assume, for $m \geq 3$, the lowest excited state
(the compound exciton)
of the ground-state Eq. (\ref{9}) as\cite{balev2006}
\begin{equation}
\Psi_{\tilde{N};(m)}^{i_{0},j_{0};\tilde{n}}= \sum_{n=-\ell}^{\ell }
C_{n}(m) \Phi_{\tilde{N},(m);n}^{i_{0},j_{0};\tilde{n}} (\mathbf{r}%
_{1},\ldots,\mathbf{r}_{\tilde{N}}) ,
\label{13}
\end{equation}
where
the excited ``partial'' many-electron wave function
$\Phi_{\tilde{N},(m);n}^{i_{0},j_{0};\tilde{n}}$ it follows
from the ground-state ``partial'' many-electron wave function
$\Psi _{\tilde{N}}^{n,(m)}$
after changing in its $\tilde{N}-$dimensional Slater determinant
of the $i_{0}-$th row $\varphi_{k_{xi_{0}}^{(n)}}^{(m)}(\mathbf{r}_{1}),
\varphi_{k_{xi_{0}}^{(n)}}^{(m)}(\mathbf{r}_{2}),\cdots,
\varphi_{k_{xi_{0}}^{(n)}}^{(m)}(\mathbf{r}_{\tilde{N}})$ by
the row
\begin{equation}
\varphi _{k_{xj_{0}}^{(n+\tilde{n})}}^{(m)}(\mathbf{r}_{1}),
\varphi_{k_{xj_{0}}^{(n+\tilde{n})}}^{(m)}(\mathbf{r}_{2}),
\cdots,\varphi_{k_{xj_{0}}^{(n+\tilde{n})}}^{(m)}(\mathbf{r}_{\tilde{N}}),
\label{14}
\end{equation}%
where $\tilde{n}=\pm 1,\ldots, \pm \ell$,
the implicit spin wave function
is omitted.
We assume
that the $i_{0}$-th unit cell defined by $n_{ys}^{(i_{0})}$ (where $m$
quasihole excitations appear that constitute the compound hole) as well as
the $j_{0}-$th unit cell (where $m$ quasielectron excitations are
mainly localized that constitute the excited
compound electron) there are well inside of MS. For the compound
exciton Eq. (\ref{13}) the spin of the compound electron is
not changed in a process of the excitation; excited states Eq. (\ref{13})
are orthonormal and they are orthogonal to the ground-state Eq. (\ref{9}).
The total charge of any quasielectron (quasihole) within MS
is given as $e/m$ ($-e/m$). Due to periodic boundary conditions
(in particular, the periodicity of single-electron wave functions)
it is clear that the charge density of the compound exciton has its
``images'' along $x$, outside MS, with the period $L_{x}^{\square}$.

The study shows that at $m \geq 3$ the lowest excited state,
Eq. (\ref{13}), of the compound exciton
(here it defines the activation gap
$E_{ac}^{(m)}$) is obtained for $n_{ys}^{(j_{0})}=
n_{ys}^{(i_{0})}$ and $\tilde{n}=\pm 1$.
We obtain that dimensionless activation gap
of the compound exciton $\Delta_{ac}^{(m)}=E_{ac}^{(m)}/(e^{2}/\varepsilon
\ell_{0}) $ is given, for $m=3, \; 5$, as
$\Delta_{ac}^{(3)} \approx 0.1016$,
$\Delta_{ac}^{(5)} \approx 0.0257$. Further, the study
shows \cite{balev2006}
that at $m=1$ the compound spin-exciton (with the change of the spin
of the excited compound electron) gives the activation gap as
$E_{ac}^{(1)}=|g_{0}| \mu_{B}B+1.1843 e^{2}/\varepsilon \ell_{0}$,
where many-body contribution is a bit smaller than relevant result
of Hartree-Fock approximation $\sqrt{\pi/2}(e^{2}/\varepsilon \ell_{0}$);
$g_{0}$ is the bare g-factor.

Assuming that the Fermi level is located within the finite energy
gap between the ground-state, Eq. (\ref{9}), and excited-states,
we calculate \cite{balev2006} (in particular, from
the Kubo formula) that the Hall conductance
$\sigma_{H}=-\sigma_{xy}=e^{2}/(2m\pi\hbar)$. I.e., for $m=3, 5, \dots$
the ground-state Eq. (\ref{9}) corresponds to the fractional Hall
effect, $\nu=1/m$. As the activation gap is experimentally observable
from the activation behavior of the direct current
magnetotransport coefficients, it is
given by the excitation gap of relevant compound exciton,
at $m \geq 3$, and by the gap of pertinent
compound spin-exciton, at $m=1$.

Point out that present study has some important similarities with
well known theory by Yoshioka, Halperin and Lee \cite{yoshioka1983}
of ground-state of the $\nu=1/3$ fractional quantum Hall
effect (for related recent works see, e.g.,
\cite{tapash1990}), using periodic boundary condition along $x$ and $y$.
In particular,  the Hamiltonian of \cite{yoshioka1983}, for $n$ electrons
in the rectangular cell $b \times a$, can be related with the
Hamiltonian, Eq. (\ref{4}),
$\hat{H}_{\tilde{N}}(\textbf{r}_{1},\ldots,\textbf{r}_{\tilde{N}})$
for $\tilde{N} \equiv n \rightarrow \infty$ electrons within MS. If to take
into account that the role of $b$ ($a$) now is played by $L_{x}^{\square}$
($L_{y}$), etc.; some minor differences are related with another form
of the Landau gauge in Ref. \cite{yoshioka1983}.
Present study shows that proper periodic boundary condition
can be totally relevant to symmetry, periodicity, correlations, and etc.
properties of a sought state; i.e., it will not
lead to any oversimplification.

\end{document}